\documentstyle{mn}

\newcommand{\ie}{{i.e.\,}}
\newcommand{\eg}{{e.g.\,}}
\newcommand{\cf}{{c.f.\,}}
\newcommand{\etal}{{\it et al.\,}}

\newcommand{\kev}{\,\mbox{${\,\rm keV}$}\,}
\newcommand{\msunyr}{\mbox{\,$M_\odot~{\rm yr}^{-1}$\,}}
\newcommand{\lsun}{\mbox{\,$L_\odot$\,}}
\newcommand{\msun}{\mbox{\,$M_\odot$\,}}
\newcommand{\ergs}{\mbox{${\,\rm erg~s}^{-1}$\,}}
\newcommand{\cmn}{\mbox{${\,\rm cm}^{-2}$\,}}
\newcommand{\kt}{\mbox{$kT$}}
\newcommand{\nh}{\mbox{$N_H$}}
\newcommand{\rosat}{\mbox{\sl ROSAT}}
\newcommand{\asca}{\mbox{\sl ASCA}}
\newcommand{\einstein}{\mbox{\sl EINSTEIN}}
\newcommand{\bllac}{{1E~0331.3-3629}}

\title[Radio and X-ray Observations of NGC\,1365]{The complex radio and
X-ray structure in the nuclear regions of the active galaxy NGC\,1365}

\author[I.\ R.\ Stevens \etal]
{Ian\ R. Stevens$^1$, Duncan\ A.\ Forbes$^1$, Ray\ P.\ Norris$^2$\\ 
$^1$ School of Physics and Astronomy, University of Birmingham, 
Edgbaston, Birmingham, B15 2TT, UK\\
$^2$ Australia Telescope National Facility, CSIRO, PO Box 76,
Epping, NSW 2121, Australia\\
irs@star.sr.bham.ac.uk, forbes@star.sr.bham.ac.uk, rnorris@atnf.csiro.au}

\date{Accepted .....................; Received .....................; 
in original form .......................}

\begin{document}

\maketitle

\begin{abstract}

We present a multiwavelength analysis of the prominent active
galaxy NGC\,1365, in particular looking at the radio and X-ray properties of
the central regions of the galaxy.

We analyse \rosat\ (PSPC and HRI) observations of NGC\,1365, and discuss
recent \asca\ results. In addition to a number of point sources in the
vicinity of NGC\,1365, we find a region of extended X-ray emission
extending along the central bar of the galaxy, combined with an emission
peak near the centre of the galaxy. This central X-ray emission is
centred on the optical/radio nucleus, but is spatially extended. The
X-ray spectrum can be well fitted by a thermal plasma model, with a
temperature of $\kt=0.6 - 0.8\kev$ and a very low local absorbing column.
The thermal spectrum is suggestive of starburst emission rather than
emission from a central black-hole.

The ATCA radio observations show a number of hotspots, located in a ring
around a weak radio nucleus. Synchrotron emission from electrons accelerated
by supernovae and supernova remnants (SNRs) is the likely origin of these
hotspots (Condon \& Yin 1990; Forbes \& Norris 1998). The radio nucleus has a
steep spectrum, indicative perhaps of an AGN or SNRs. The evidence for a jet
emanating from the nucleus (as has been previously claimed) is at best
marginal. The extent of the radio ring is comparable to the extended central
X-ray source.

We discuss the nature of the central activity in NGC\,1365 in the light of
these observations. The extended X-ray emission and the thermal spectra
strongly suggest that at soft X-ray energies we are not seeing emission
predominantly from a central black-hole, although the presence of Fe-K
line emission at higher energies does suggest the presence of an AGN.
Consequently, a black-hole is probably not the dominant contributor to
the energetics of the central regions of NGC\,1365 at radio, optical or
soft X-ray wavelengths. Activity associated with a starburst is likely the
dominant explanation for the observed properties of NGC\,1365.

\end{abstract}


\section{Introduction}

It is becoming clear that in Seyfert galaxies there is a close link between
the Seyfert nucleus, which may harbour a super-massive black-hole, and
circumnuclear starburst activity. The nature of the connection is unclear,
whether the starburst is somehow responsible for the formation or perhaps
fuelling of a central black-hole, or whether activity associated with the
central black-hole is responsible for triggering the star-formation. To
understand the complexities of the central regions of active galaxies and the
relationship between Seyfert and starburst activity we need detailed
multiwavelength studies of individual galaxies. In this paper we present such
a study of the prominent spiral galaxy NGC\,1365, concentrating on X-ray and
radio data.

NGC\,1365 is classified as a SBb(s)I type barred spiral galaxy by Sandage \&
Tammann (1981), and is a member of the Fornax cluster. NGC\,1365 is an active
galaxy with an emission line nucleus (with both broad and narrow H$\alpha$
emission) located at the edge of a dust-lane, which runs across the central
bulge (V\'{e}ron \etal 1980). We adopt a distance of 20\,Mpc for NGC\,1365
throughout this paper.

The active nucleus of NGC\,1365 has been multiply classified. For instance, on
the basis of the FWHM of the H$\alpha$ component Hjelm \& Lindblad (1996)
classified NGC\,1365 as a Seyfert 1.5 (see also Kristen \etal 1997). Turner,
Urry \& Mushotzky (1993) classified it as a Seyfert 2, and Maiolino \& Rieke
(1995) as a Seyfert~1.8.

In addition to the active nucleus, NGC\,1365 contains a large number of
giant H{\small II} regions, with several of them in close proximity to
the nucleus (Alloin \etal 1981). {\sl HST} Faint Object Camera
observations of NGC\,1365, discussed by Kristen \etal (1997), reveal
numerous bright super star-clusters around the nucleus. These clusters
are very compact, with radii estimated to be $\leq 3$pc, and tend to fall
on an elongated ring around the nucleus. Correlations between the
locations of these super-star clusters and sources in this paper will be
discussed later.

Hjelm \& Lindblad (1996) have also discussed optical observations of the
central regions of NGC\,1365, and in particular the conical region of high
excitation gas associated with the nucleus. They describe a kinematical model
of this region, whereby gas is accelerated in a bi-conical region, with the
cone axes aligned with the galaxy rotation axis. The cone is also aligned with
a suggested radio jet (Sandqvist, J\"{o}rs\"{a}ter \& Lindblad 1995), though
we shall discuss the nature of this jet in more detail later. Radio
observations of NGC\,1365 by Saikia \etal (1994), Sandqvist \etal (1995) and
Forbes \& Norris (1998) reveal a complex, possibly ringlike structure around
the nucleus, but with several radio hotspots loosely coincident with the giant
H{\small II} regions.

One of the nuclear H{\small II} regions (region L4 in the terminology of
Alloin \etal 1981) shows a broad emission line at around 5696\AA, due to
C{\small III} emission (Phillips \& Conti 1992). The strength of this line,
and the absence of C{\small IV} 5808\AA\ is indicative of the presence of
between 350 and 1400 WC9 Wolf-Rayet (WR) stars. The presence of such a
substantial population of massive WR stars is indicative of very recent
($3-6$\,Myr) star-formation close to the nucleus. NGC\,1365 was not included in
the original WR galaxy catalogue of Conti (1991), but was included in the
subclass of WR barred spiral galaxies by Contini, Davoust \& Consid\`{e}re
(1995). We shall discuss the properties of NGC\,1365 in relation to WR galaxies
in Section~4.1.

The nuclear regions of NGC\,1365 were seen as a moderately strong IR source
with {\sl IRAS} (Ghosh \etal 1993). NGC\,1365 was also observed as a soft X-ray
source by the \einstein\ satellite, with an X-ray luminosity of $L_x=8.2\times
10^{40}\ergs$ (for the $0.2-4.0\kev$ waveband, corrected for distance;
Fabbiano, Kim \& Trinchieri 1992). Komossa \& Schulz (1998) have also
recently presented a study of the X-ray properties of NGC\,1365.

In this paper we present a detailed analysis of radio and X-ray observations
of NGC\,1365 in an attempt to better understand the puzzling central regions of
this galaxy. The X-ray data probe the hot thermal gas from starburst emission
and the hard non-thermal radiation from any massive compact object. Radio
observations on the other hand allows us to look for non-thermal emission from
synchrotron radiation from SNRs or an AGN, or for thermal free-free emission
from young stellar populations of H{\small II} regions. The combined
X-ray/radio morphology will in turn allow us to to investigate the origin of
the emission.

NGC\,1365 has been observed on several occasions at X-ray energies. Here
we shall concentrate on observations with the \rosat\ satellite (both
PSPC and HRI instruments), but also discuss recent results from the
\asca\ satellite. These satellites and instruments have a range of
complementary properties - the \rosat\ PSPC is very sensitive, has
reasonable spectral resolution and reasonable spatial resolution ($\sim
25^{''}$), allowing us to probe the spectral properties of central
regions and look for any extended emission; the \rosat\ HRI is less
sensitive and has essentially no spectral resolution, but has excellent
spatial resolution ($\sim 5^{''}$) allowing a detailed study of the X-ray
morphology of NGC\,1365.
\asca\ on the other hand has poor spatial resolution (a few arcminutes), but
excellent spectral resolution, which coupled with a broader bandpass
($0.5-10\kev$, as compared to $0.1-2.4\kev$ for \rosat) allows us to
investigate the spectral properties of the integrated emission from
NGC\,1365 in more detail. In particular, \asca\ allows us to look for
harder X-ray emission that might be associated with a Seyfert nucleus.
Radio observations from the VLA have been previously reported (\eg Sandqvist
\etal 1995). Here we use the Australia Telescope Compact Array (ATCA)
observations of Forbes \& Norris (1998) which gives a more circular beam than
the VLA for a galaxy at Dec. $\sim -36^{\circ}$. We report on observations
at both 3cm and 6cm.

Some of the important issues that we intend to address in this paper are as
follows:

\begin{enumerate}

\item What is the origin of the X-ray emission from NGC\,1365; \ie is it
associated with a massive central black-hole or is it due to starburst
activity?

\item What is the origin of the radio hotspot emission?

\item What is the energetic importance of the Seyfert nucleus?

\item Do we find any complimentary evidence for jets, winds, outflows or
emission line cones at radio/X-ray frequencies?

\end{enumerate}

The paper is organised as follows; in Section~2 we discuss the X-ray
observations of NGC\,1365 with the \rosat\ and \asca\ satellites, in Section~3
we discuss the ATCA radio data, and in Section~4 we discuss what the radio and
X-ray data tell us about the nature of the central activity in NGC\,1365.

\begin{table*}
\caption{A summary of the X-ray and radio observations of NGC\,1365 
discussed in
this paper}
\begin{center}
\begin{tabular}{lccc}
\\
Instrument & Date & PI & Duration\\
\\
\multicolumn{2}{l}{X-ray Observations}  & &\\
\rosat/PSPC     & 1992 Aug 22 -- Aug 23  & T.\,J. Turner  & 2.6\,ksec  \\
\rosat/PSPC     & 1993 Feb 05 -- Feb 10  & T.\,J. Turner  & 6.2\,ksec  \\
\rosat/HRI      & 1994 Jul 20 -- Aug 04  & K.\,P. Singh   & 9.9\,ksec  \\
\rosat/HRI      & 1995 Jul 03 -- Jul 04  & K.\,P. Singh   & 9.8\,ksec  \\
&&&\\
\multicolumn{2}{l}{Radio Observations}  & &\\
ATCA/6cm        & 1994 Nov 10 -- Nov 11  & D.\,A. Forbes & 36\,ksec \\
ATCA/3cm        & 1994 Nov 10 -- Nov 11  & D.\,A. Forbes & 36\,ksec \\
\end{tabular}
\end{center}
\label{tabx1}
\end{table*}

\section{X-ray Observations of NGC\,1365}

A summary of the X-ray observations discussed in this paper is given in
Table~\ref{tabx1}. In all cases we create background subtracted images,
and use the point source searching (PSS) package (Allan 1993) to search
for unresolved sources. Turner \etal (1993) have discussed the point
sources in the vicinity of NGC\,1365 detected with the \rosat\ PSPC and
suggest some plausible candidates, namely X-ray binaries, H{\small II}
regions in the spiral arms and possibly background AGN (see also Komossa 
\& Schulz 1998). Several of the
candidates are embedded in the disk of the galaxy and it is difficult to
identify counterparts. Because we have also detected diffuse X-ray
emission there is also the possibility that some of the HRI sources are
merely spurious detections of irregularities in the diffuse emission.

For the \rosat\ PSPC observations we extracted source spectra for regions
around NGC\,1365 and fitted them with simple spectral models. More details
are given in the appropriate sections below.

The $z=0.308$ BL~Lac object \bllac\ is in the field of view of all the
observations. This X-ray source was discovered by the \einstein\ {\it Extended
Medium Sensitivity Survey}. The optical identification is discussed in
Maccacaro \etal (1994). The X-ray properties of \bllac\ are discussed in
Lamer, Brunner \& Staubert (1996). We have cross checked the position of
\bllac\ with the APM catalogue (Irwin 1992), and the optical position is well
determined to arcsecond accuracy. We adopt a position of ${\rm RA}=03\ 33\
12.3, {\rm Dec.} = -36\ 19\ 47$ (J2000) for \bllac\ (\cf Maccacaro \etal
1994). Consequently we can use the X-ray determined position of \bllac\ to
accurately align the X-ray images. This is important as it enables us to
accurately define the location of the X-ray emission in NGC\,1365, and relate
it to the radio emission, and hence understand its origin.

\begin{figure}
\vspace{9cm}
\includegraphics{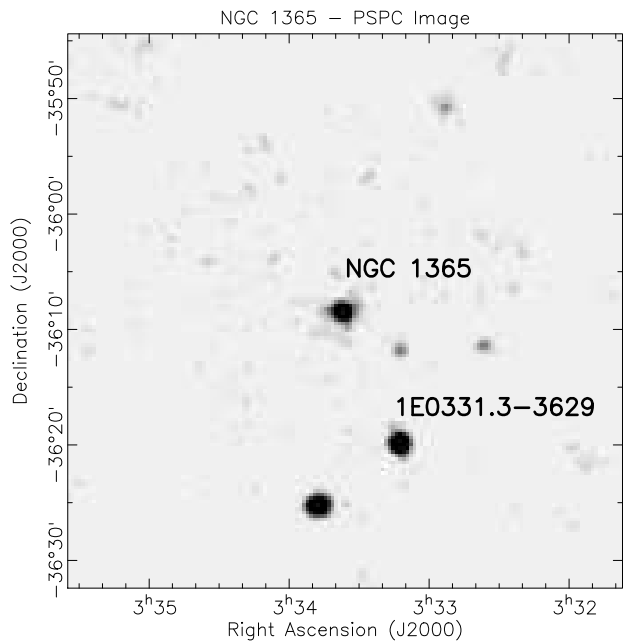}
\caption{The central $0.8\times 0.8$ degree region of the 
\rosat\ PSPC field containing NGC\,1365. This is the longer, 1993 Feb,
observation, with an observation length of 6.2\,ksec. The X-ray image has been
background subtracted, and has been extracted with a $5^{''}$ pixel size and
smoothed with a Gaussian of FWHM $10^{''}$. 
There is strong X-ray emission from NGC\,1365, at 
the centre of the field, along
with several other point sources. The strong source to the SW of NGC\,1365 is
the BL~Lac object \bllac, which is apparent in all the X-ray
observations and is used to register the X-ray images.
The source to the SE of NGC\,1365 is an unclassified 
object, possibly associated
with a faint optical object from the Digital Sky Survey, which is strongly
variable, and is extremely weak in the later HRI observations. The detector 
ring structure of the PSPC is barely visible.}
\label{figx1}
\end{figure}

\begin{figure*}
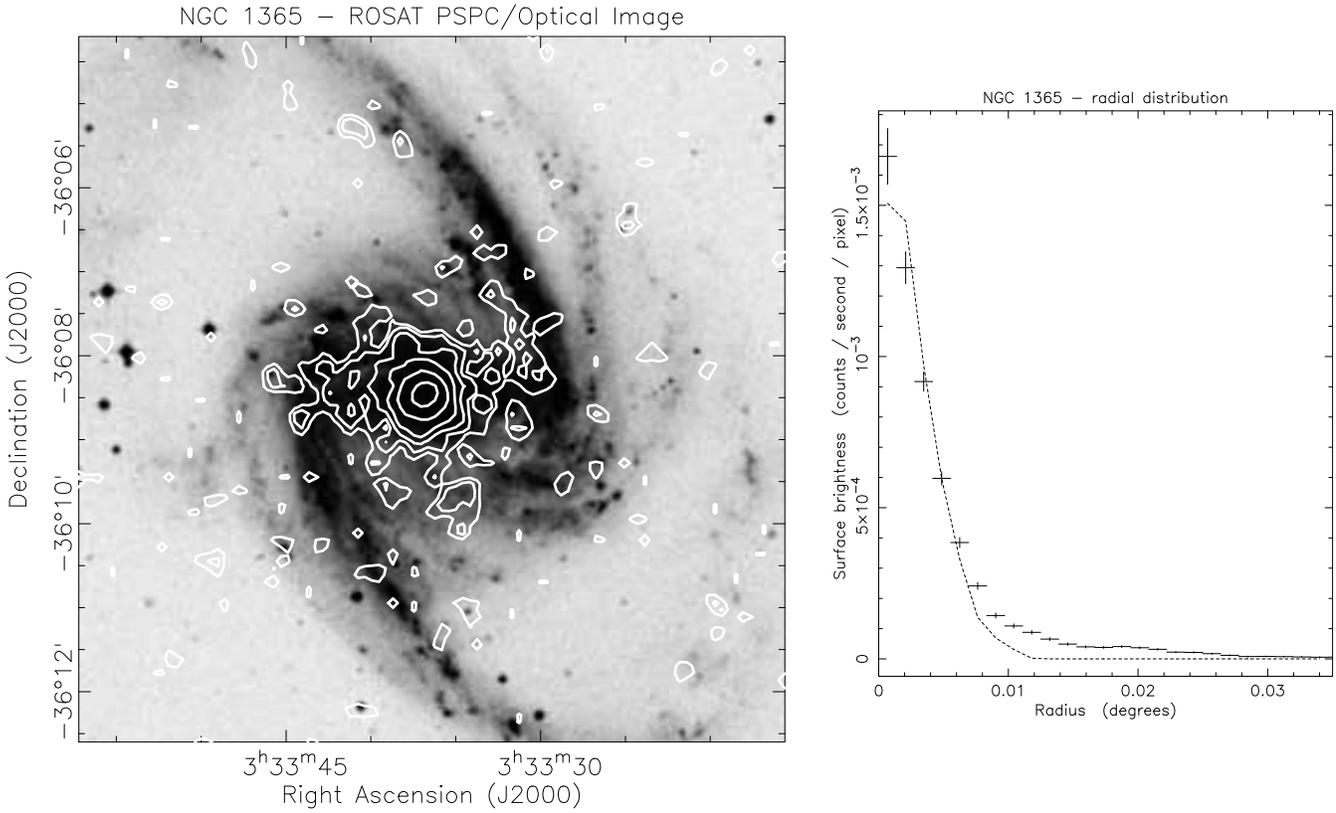

\vspace{12cm}
\includegraphics{n1365_pic2a.ps}
\includegraphics{n1365_pic2b.ps}
\caption{Left panel: The X-ray emission from NGC\,1365 as seen with the 
\rosat\ PSPC. The X-ray contours from the longer PSPC observation are
superimposed on the Digitized Sky Survey image of NGC\,1365. The lowest contour
is at a level of $5.6\times 10^{-3}$ cts s$^{-1}$ arcmin$^{-2}$, and increase
by a factor 2. Clearly visible is strong X-ray emission from the central
region of NGC\,1365, as well as extended emission along the galaxy bar,
extending out to a radius of about 1.5~arcmin. Several other point sources are
also visible and are discussed in Section~4. Right panel: The radial profile
of the X-ray emission from the PSPC observations of NGC\,1365.
Also shown is the PSF for the \rosat\ PSPC assuming a
spectrum with a mean photon energy of $0.7\kev$. The X-ray emission is clearly
extended, with a radius of about 1.5~arcmin, but also with a point-like
central source.}
\label{figx2}
\end{figure*}

\subsection{\rosat\ PSPC Observations}

NGC\,1365 has been observed with the \rosat\ PSPC on two occasions (with an
observation length of 2.6\,ksec for observation No.1, and 6.2\,ksec for
observation No.2, with the observations separated by about 6 months). Results
from these observations have been reported by Turner \etal (1993), who also
reported on PSPC observations of 5 other Seyfert~2 galaxies.

Turner \etal (1993) found a total of 5 point sources near the nuclear source
in NGC\,1365, and found that the emission from the nuclear source was not well
fitted by an absorbed power-law, but could be better fitted by either a model
with a power-law, plus a Raymond-Smith thermal component (with $\kt=0.6\kev$),
or by a power-law with an emission line at an energy of $0.8\kev$. It is worth
noting that the fit quality for the other Seyfert galaxies, assuming a
power-law model, was substantially better than for NGC\,1365.

We have reanalysed both \rosat\ PSPC observations. We extracted background
subtracted spectral images for both observations. We adjusted the pointing of
both observations using the BL~Lac object \bllac. The X-ray image
(Figure~\ref{figx1}) shows moderately strong X-ray emission from 
NGC\,1365 (the precise location will be discussed in Section~4.2) as well 
as emission from other
sources near the galaxy. The strong source to the SW of NGC\,1365 is
\bllac. 

In Figure~\ref{figx2}, for the longer (6.2\,ksec) observation, we show
X-ray contours of the emission from NGC\,1365 superimposed on an optical
image from the Digitized Sky Survey. We also show in Figure~\ref{figx2}
the radial profile for the X-ray emission, compared to the Point Spread
Function (PSF) for the PSPC, for a spectrum with a mean photon energy of
$0.7\kev$. The X-ray emission is clearly extended well beyond the PSF of
the \rosat\ PSPC. Also, the extended emission is not azimuthally
symmetric, with low surface brightness emission stretching preferentially
along the bar of the galaxy. There is also point-like X-ray emission from
the central regions of the galaxy. However, as we will discuss later in
regard to the HRI observations, this source, which is centred on the
optical/radio nucleus, is not point-like, but is extended. Also
visible is a point source in the disk of the galaxy to the SW of the
nucleus. This source is discussed in more detail in 
Komossa \& Schulz (1998).

\begin{figure*}
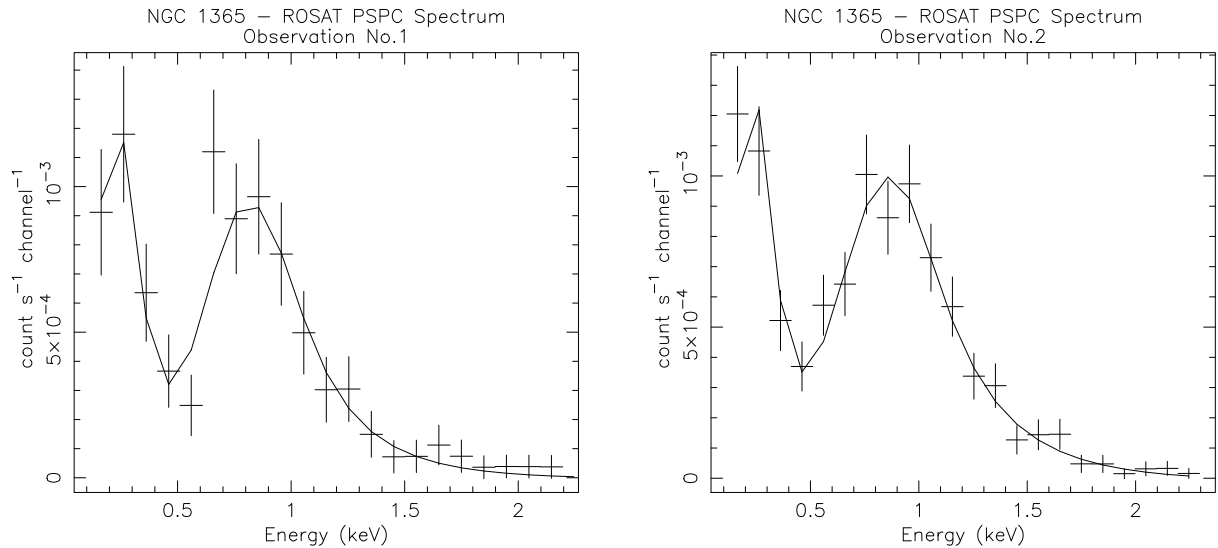

\vspace{8cm}
\includegraphics{n1365_pic3a.ps}
\includegraphics{n1365_pic3b.ps}
\caption{The \rosat\ PSPC spectra of NGC\,1365 for both observations. The
spectra have been fitted independently and the  results are given in 
Table~\ref{tabx2}. The model used here is a single temperature Raymond-Smith
plasma  model.}
\label{figx3}
\end{figure*}

In order to investigate the spectral properties of the X-ray emission we have
extracted a spectrum from the region within 1.5~arcmin of the central point
source for both PSPC observations, and we fit them independently to look for
any variability.

We fit both spectra with two different spectral models, i) an absorbed
power-law and ii) a single temperature Raymond-Smith plasma model. The
details of the fits are given in Table~\ref{tabx2}. We find that the
power-law model gives a poor fit (as was found by Turner \etal 1993),
with reduced $\chi^2$ of 1.5 and 2 respectively for the two observations.
On the other hand, for the Raymond-Smith plasma models we get good fits
(reduced $\chi^2\leq 0.7$), with $\kt\sim 0.6 - 0.8\kev$, a low
metallicity and a column ($\nh\sim 2\times 10^{20}\cmn$).

The derived luminosities in the \rosat\ waveband ($0.1-2.5\kev$), 
corrected for absorption, are $5 -6\times 10^{40}\ergs$. Adding in an
absorbed power-law component to a thermal model (as might be
expected if we had a highly absorbed nuclear source) does not lead to a
better fit with the PSPC (\cf Section~2.3).

For the absorbed thermal spectral model, there are some differences
in the fitted parameters, mostly notably in the temperature. The
differences are only significant at the $1-2\sigma$ level and may not
reflect any physical changes.

Komossa \& Schulz (1998) also present results from an analysis of 
the \rosat\ PSPC observations of NGC\,1365. They too note that a 
single temperature Raymond-Smith model provides a successful fit 
to the spectra. However, their preferred model is for a two 
component model, consisting of a Raymond-Smith model plus a powerlaw. 
The main reasons that 
Komossa \& Schulz (1998) give for preferring the two component model is that 
the best-fit single temperature Raymond-Smith model has a low abundance
($Z\sim 0.1Z_\odot$). This low abundance is at odds with optical 
abundance determinations (Alloin \etal 1981)
However, we note that the work of Strickland \& Stevens 
(1998) suggests fitting simple (one or two component) 
spectral models to intrinsically multi-temperature gas 
(such as would be expected in a starburst) can result 
in major errors in the fitted metallicity. Consequently, 
the discrepancy between optical and X-ray determinations 
should perhaps not be taken too seriously. 
Komossa \& Schulz (1998)  also note that a Raymond-Smith plus 
power-law model is more in line with the \asca\ results of Iyomoto \etal 
(1997 - see below). On the basis of our spectral fitting of the 
PSPC spectra we find no statistical reason to prefer a two 
component model over our single temperature Raymond-Smith model.

It is notable that the X-ray properties of NGC\,1365, as observed by the
PSPC, are similar, in general terms, to those of starburst galaxies (see
Section~4), both in terms of showing extended emission and having an
X-ray spectra better fit with a thermal model than a power-law.
Also, the fitted column density is very similar to the Stark value
($\nh=1.92\times 10^{20}\cmn$, Stark \etal 1992)  
It is clear that we are not seeing a
significant amount of emission in the PSPC waveband from a heavily
absorbed Seyfert nucleus, However, the power-law component in such
systems (for example, NGC\,1386, Iyomoto \etal 1997; NGC\,7582, Xue \etal
1998), associated with the nucleus, has a column $\geq 10^{23}\cmn$,
which means that it will not be detected in the PSPC waveband. We also
note that in the case of NGC\,1068, a Seyfert~2 galaxy, the power-law
component in the X-ray spectrum has a low column (Ueno \etal 1994). In
NGC\,1068 the interpretation is that we are seeing scattered emission from
the nucleus, rather than direct emission, the column being too high for
it to be directly observed.

We shall discuss the origin of the X-ray emission and the relation to
the emission from other Seyfert galaxies in Section~4.

\begin{table*}
\caption{Spectral analysis of \rosat\ PSPC observations of 
NGC\,1365} 
\begin{center}
\begin{tabular}{lccccccc}
\\
& \multicolumn{3}{c}{Raymond-Smith} & \multicolumn{2}{c}{Power-law} & &  \\
& \multicolumn{3}{c}{spectral model } & \multicolumn{2}{c}{spectral model} & &  \\
& \multicolumn{3}{c}{\hrulefill} & \multicolumn{2}{c}{\hrulefill} & &  \\
Spectral & \nh     & $kT$ & $Z$       & $N_H$ &$\alpha$&$\chi_\nu^2$& $L_x$ \\
Model  &($10^{20}\cmn$)&(keV) &($Z_\odot$)&($10^{20}\cmn$)&&&(erg~s$^{-1}$) \\
(1)   & (2)& (3) & (4) & (5) & (6) & (7) & (8)\\
PSPC Observation No. 1 & & & & &\\
AG*RS & $1.78^{+0.65}_{-0.65}$ & $0.58^{+0.09}_{-0.09}$ & 
$0.074^{+0.05}_{-0.04}$ & -- &--   & 0.69 (17) & $5.1\times 10^{40}$ \\
AG*PL &  --  & --   & --   & $5.40^{+1.1}_{-1.1}$ &
$3.01^{+0.32}_{-0.30}$ & 1.47 (18) & $2.1\times 10^{41}$ \\
 & & & & &\\
PSPC Observation No. 2 & & & & &\\
AG*RS & $1.91^{+0.45}_{-0.45}$ & $0.77^{+0.07}_{-0.06}$ & 
$0.076^{+0.04}_{-0.04}$ &--&--& 0.69 (18)&$6.2\times 10^{40}$ \\
AG*PL & -- & -- & -- & $5.64^{+0.07}_{-0.07}$ & $2.89^{+0.19}_{-0.18}$ & 
2.09 (19) & $2.3\times 10^{41}$ \\ 
\end{tabular}
\end{center}
\begin{flushleft}
\noindent Notes on Table\ref{tabx2}:\\
Column 1: Spectral model used to fit the data, AG - absorbing component; 
RS - single temperature Raymond-Smith plasma model; PL - power-law model.\\
Columns 2-4: Best fit parameters (with errors) for single temperature
Raymond-Smith model.\\
Columns 5-6: Best fit parameters (with errors) for power-law spectral model.\\
Column 7: Value of the reduced $\chi^2$ (and number of degrees of freedom in
brackets) for the spectral fits.\\
Column 8: The intrinsic X-ray luminosity (\ie corrected for
absorption). The quoted luminosities for
the \rosat\ PSPC fits are for the $0.1-2.5\kev$ energy band.\\
\end{flushleft}
\label{tabx2}
\end{table*}

\subsection{\rosat\ HRI Observations}

There have been two \rosat\ HRI observations of NGC\,1365 (Table~\ref{tabx1}),
of duration 9.8 and 9.9\,ksec, separated by approximately 18 months. The HRI is
less sensitive than the PSPC for diffuse emission but is much better for
studying smaller scale spatial structure. The combination of the two
instruments allows a detailed analysis of both point source emission and
the extended diffuse emission in the central regions of NGC\,1365.

The method of analysis that we have adopted for the HRI observations is as
follows; we sort a background subtracted image using only channels $3 - 8$ to
minimise the background (Briel \etal 1995). We then search for sources using
PSS and adjust the pointing using the position of the BL~Lac object
\bllac. The centroids of the X-ray emission from the \rosat\ HRI and
PSPC, as determined by the PSS package, are given in
Table~\ref{tabx3}. We note that the positions of the X-ray centroids only
differ from the mean X-ray centroid position by $\leq 3^{''}$, and that the
mean centroid of the X-ray emission is consistent with the position of the
optical/radio nucleus. This point will be discussed in more detail in
Section~4.2.

\begin{figure*}
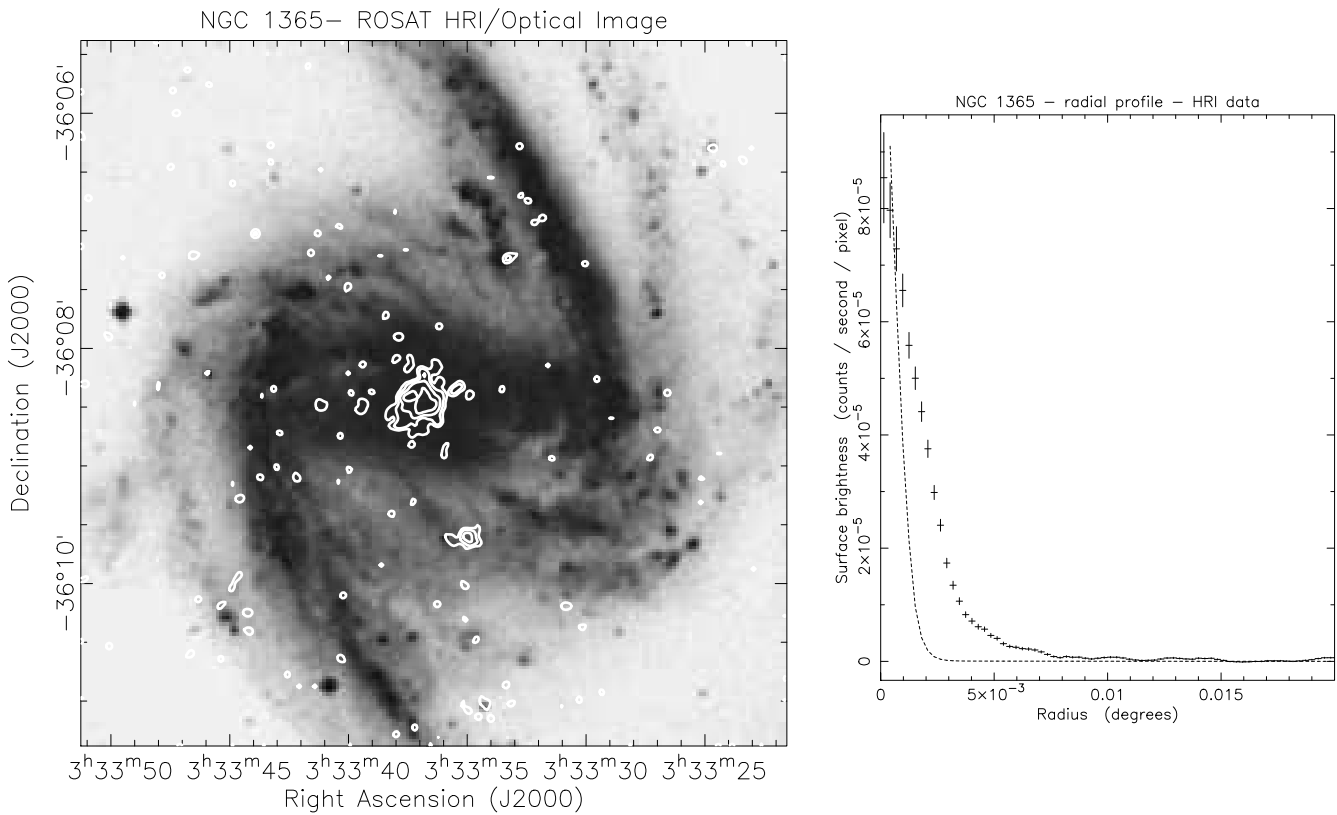

\vspace{12cm}
\includegraphics{n1365_pic4a.ps}
\includegraphics{n1365_pic4b.ps}
\caption{Left panel: The X-ray emission from NGC\,1365 as seen with
the \rosat\ HRI. The X-ray contours are from the longer HRI observation and
are superimposed on an optical image from the Digitized Sky Survey image. The
X-ray image has been background subtracted, and has been extracted with a
$1^{''}$ pixel size and smoothed with a Gaussian of FWHM $3^{''}$. The lowest
contour level is at a level of $2.2\times 10^{-2}$~cts~s$^{-1}$~arcmin$^{-2}$
and the contours increase by a factor 2. There is strong X-ray emission from
the central point source, as well as several point sources near to the nuclear
source. The BL~Lac object used to register the image is not shown. The point
source to the SW of the nucleus is probably associated with an H{\small II}
region visible in this image. The spatial resolution of the
\rosat\ HRI is $\sim 5^{''}$. Right panel: The radial 
profile of the central point
source compared to the
\rosat\ HRI PSF. The central source is clearly extended.}
\label{figx4}
\end{figure*}

We have generated background subtracted images from both HRI 
observations. The X-ray/optical morphology as seen with the longer 
HRI observation is shown in Figure~\ref{figx4}. The main features in 
the X-ray morphology
are the central source (which appears to be extended with some evidence
of substructure), along with the several other point sources in the disk
of the galaxy. Also in Figure~\ref{figx4} we show the radial profile of
the central X-ray emission, in conjunction with the PSF for the \rosat\
HRI. In the PSPC we saw low surface brightness diffuse emission, in
conjunction with point-like emission. The HRI gives us an improved
picture of the central regions. While there is also low-surface
brightness emission seen with the HRI, the central source is now seen to
be clearly resolved, with an extent of maybe $15-20^{''}$. Whether this
is due to multiple point sources or genuinely diffuse emission is
unclear, but its extent is very similar to that of the radio ring
(Section~3). Also, in the PSPC observations the emission was extended
preferentially along the bar of the galaxy. In the HRI observations we
are sampling emission from a smaller, more central region, and in
contrast to the PSPC observations there is no evidence of significant
azimuthal variation in the HRI observations.

The point source searching program PSS detected several other point sources in
close vicinity to the nucleus. In addition to genuine point sources we may
also be detecting peaks in the more extended emission seen in the PSPC. The
fact that we definitely see extended X-ray emission from the central regions
with the HRI strongly suggests that we are genuinely seeing diffuse emission.

\subsection{\asca\ Observations}

Because of its wider waveband ($0.5-10.0~\kev$) and higher spectral
resolution, \asca\ observations of NGC\,1365 can potentially tell us more
about the origin of the X-ray emission from the central source - in
particular, whether there is a highly absorbed power-law component, as is
seen in some other Seyfert~2 galaxies.

On account of the poor spatial resolution of \asca\ (typically $\sim
3^{'}$), and the presence of the source in the disk of the galaxy there
is likely to be a problem of contamination in the spectra of the central
source.

Results from \asca\ observations of NGC\,1365 have already been reported
on in some detail in Iyomoto \etal (1997). Rather than repeat the
analysis we shall quote the main results from this paper, which can be
summarised as follows:

\begin{enumerate}

\item A hard point-like source was detected at a position coincident with
the centre of NGC\,1365.

\item The continuum spectra could be fitted with a power-law (with a
photon index of $\sim 0.8$) plus a Raymond-Smith 
plasma model (with a $\kt=0.85\kev$ and a metallicity of $0.2Z_\odot$).
At energies $<2\kev$ the thermal component dominates over the power-law.

\item The power-law component did not require substantial excess absorption.

\item The $2-10\kev$ luminosity of the power-law component was found to be
$4\times 10^{40}\ergs$ (for an assumed distance of 20\,Mpc).

\item Also present in the spectrum was a strong, broad Fe-K line, with 
a line centre energy of 6.6\kev and a width of 0.2\kev.

\end{enumerate}

These results are very important for a number of different reasons.
First, the detection of strong Fe-K emission from a galaxy is usually an
indicator of an active nucleus, although the luminosity of the power-law
component in NGC\,1365 is rather low compared to the power-law components
in other Seyfert galaxies. A second important point is that the power-law
component is not highly absorbed. This is similar to NGC\,1068 (Ueno \etal
1994), but dissimilar to several other Seyfert~2 galaxies, where in
addition to a thermal component a hard {\em highly absorbed}
power-law component is seen (examples include NGC\,1386, Iyomoto \etal
1997; NGC\,4388, Iwasawa \etal 1997; NGC\,7582, Xue \etal 1998).

The metallicity for the Raymond-Smith component of the fit to the \asca\
data ($0.2Z_\odot$) is somewhat higher than the metallicity determined
from the PSPC results. However, given the respective error-bars the
results are not inconsistent (see also Strickland \& Stevens 1998). 

We shall discuss the implications of the \asca\ results at length in
Section~4.

\begin{table}
\caption{The centroids of the X-ray emission from the central source in
NGC\,1365 as compared to the positions of the optical and radio position of the
nucleus (Alloin \etal 1981; Forbes \& Norris 1998)} 
\begin{center}
\begin{tabular}{lcc}
Observation  & RA     & Dec \\
             &\multicolumn{2}{c}{(J2000)}  \\
PSPC Obs. 1 & 03 33 36.5 & $-$36 08 25 \\
PSPC Obs. 2 & 03 33 36.4 & $-$36 08 27 \\
HRI Obs. 1  & 03 33 36.4 & $-$36 08 28  \\
HRI Obs. 2  & 03 33 36.4 & $-$36 08 25 \\
\\
Average     & 03 33 36.4 & $-$36 08 26.3 \\
\\
Optical Position & 03 33 36.4 & $-$36 08 25.7  \\
Radio Position & 03 33 36.35 & $-$36 08 25.9  \\
\end{tabular}
\end{center}
\label{tabx3}
\end{table}

\section{Radio Observations of NGC\,1365}

\begin{figure*}
\vspace{10cm}
\includegraphics{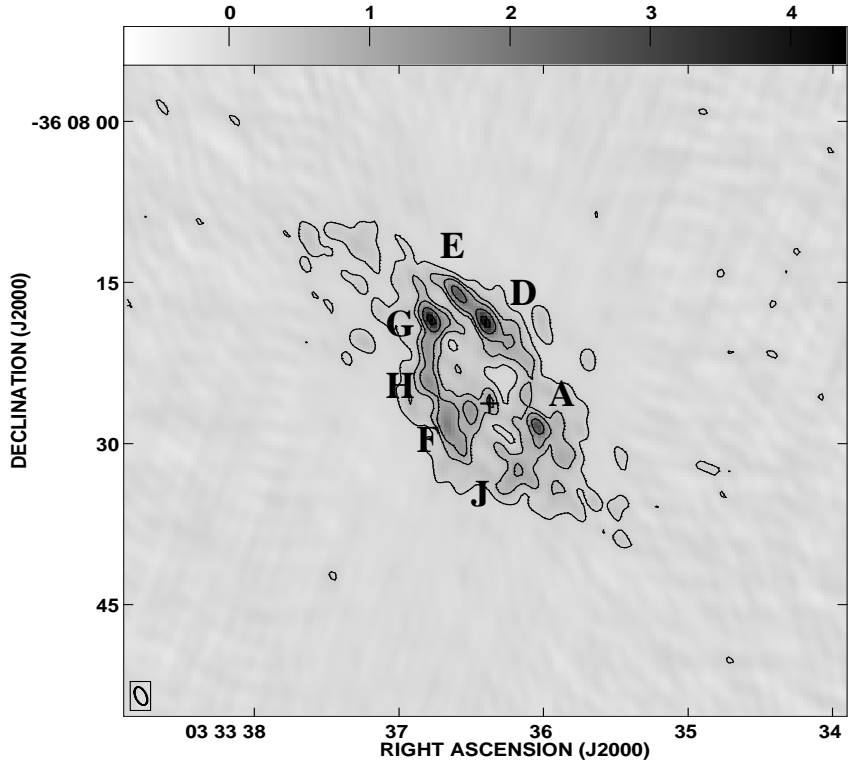}
\caption{The 3cm ATCA radio map of the central region of NGC\,1365.
A number of hotspots in an elongated ring structure surrounding a weak nucleus
can be seen, and are labelled as in Table~4. The position of the nucleus is
marked with a cross. The beam size ($\sim 1^{''}$) is shown in the lower left.
The upper greyscale bar shows the intensity scale in mJy/beam.}
\label{figx7}
\end{figure*}

Radio continuum observations of NGC\,1365 using the ATCA have been presented by
Forbes \& Norris (1998). Radio maps were obtained at 3cm and 6cm giving an
effective resolution of $\sim 1^{''}$ and $\sim 2^{''}$ respectively. Forbes
\& Norris (1998) examined the overall radio properties of NGC\,1365
along with five other southern active galaxies. They did not discuss the
individual hotspots in any detail. In Figure~\ref{figx7} we show only the 3cm
radio maps of the central regions. 

The radio maps reveal a weak nucleus surrounded by a elongated $\sim 8^{''}
\times 20^{''}$ (a/b = 0.4) ring of hotspots. Sandqvist \etal (1995) 
labelled the
hotspots A--H, although B is blended with A, and C is blended with D at our
$1^{''}$ resolution. Forbes \& Norris (1998) identified an additional hotspot
to the SW, called `J'. Components E, H and J are marginally resolved in our
3cm map, so that typical component sizes are $\sim 1^{''}$ or 100~pc at our
assumed distance of 20\,Mpc. Centering on the component flux peak, we have
measured positions and $2^{''}$ diameter fluxes. We have chosen to subtract
the background in an annulus around each component rather than subtract a
constant background as done by Sandqvist \etal (1995). Thus we do not expect
very close agreement between our measurements (given in Table~\ref{tabx4})
and those of Sandqvist \etal (1995). However we might 
expect our 6/3 cm spectral
index to be similar to their 6/2~cm index. For the nucleus itself Sandqvist
\etal (1995) 
find $\alpha$ $< -0.87$ for 6/2~cm, which is consistent with our value
of $\alpha = -1.1$ for 6/3 cm. For components A, D and F, we get comparable
indices. In the case of component E and G, their values (ie $\alpha = -0.77$
and $-0.44$ respectively) are considerably different from ours
($\alpha = -0.4$ and $-0.04$). Most of the
components in the ring have moderate to steep spectral indices which are
consistent with non-thermal emission. The typical hotspot 6cm luminosity
is 10$^{36}$ erg s$^{-1}$ which, along with the steep spectral index, suggests
that the radio emission from each hotspot is made up of several SNRs.

Sandqvist \etal (1995) have suggested that component F represents the end of a
\lq\lq jet-like feature\rq\rq\ emanating from the 
nucleus. They found component F to
have the steepest spectral index, other than the nucleus. Our data indicate
that both component H and J have similarly steep indices. An alternative
explanation to the radio jet is that of a nuclear bar which extends out to the
radio ring. Enhanced star formation is often seen at the ends of a bar in
barred galaxies. The nature of the nucleus in NGC\,1365 is difficult to confirm
with our radio data. It is unresolved in the 3cm map. The steep spectral
index indicates non-thermal synchrotron from either an AGN or SNRs. We note
that energetically a small number of Cas~A-like SNRs can account for the radio
emission. 

\begin{table}
\caption{NGC\,1365 Radio Hotspots} 

\begin{center}
\begin{tabular}{lcccc}
Hotspot & Position   & 6cm flux & 3cm flux & Sp. Index \\
        & (s,$^{''}$) & (mJy) & (mJy) & \\
Nuc. & 36.35, 25.9 & 0.46 & 0.24 & --1.1\\
A    & 36.02, 28.3 & 1.25 & 0.97 & --0.4\\  
D    & 36.37, 18.6 & 2.11 & 1.47 & --0.6\\
E    & 36.57, 16.0 & 1.01 & 0.80 & --0.4\\
F    & 36.63, 28.0 & 1.06 & 0.63 & --1.0\\
G    & 36.75, 18.3 & 1.60 & 1.56 & --0.04\\
H    & 36.76, 24.1 & 1.00 & 0.58 & --1.0\\
J    & 36.15, 32.8 & 0.65 & 0.37 & --1.0\\
\end{tabular}
\label{tabx4}
\end{center}
\begin{flushleft}
\noindent Notes to Table~\ref{tabx4}.\\
\noindent Galaxy nucleus at $03^{h}33^{m}36.35^{s}$, 
$-36^{\circ}08^{'}25.9^{''}$ (J2000).
\end{flushleft}
\end{table}

\section{Discussion}

The main points that can be distilled from these observations are as follows:

\begin{enumerate}

\item The \rosat\ PSPC observations show extended X-ray emission from
the central region of NGC\,1365, with the emission preferentially 
extended along the bar of the galaxy.
The PSPC spectra can be well fitted with a thermal
model with $\kt\sim 0.6-0.8\kev$ and a low absorbing column, comparable to
the Galactic Stark value.

\item The \rosat\ HRI observations show that the more 
concentrated central X-ray emission is still extended, with a radius of $\sim
15^{''}$ (1.5\,kpc), similar in extent to the radio ring.

\item The centroid of the X-ray emission is consistent with the position of the
radio/optical nucleus (see Section~4.2).

\item The \asca\ observations suggest that, in addition to the softer thermal
emission, there is also a power-law component which is not highly absorbed.

\item The \asca\ results also show strong Fe-K emission, indicative of
an active nucleus.

\item The radio observations reveal a steep spectrum but the radio emission
from the nucleus is less than that of the surrounding hotspots. The
evidence for a radio jet is marginal. 
\end{enumerate}

\subsection{Broad-band properties}

Having estimated the total X-ray emission from NGC\,1365 (\ie $L_x \sim 6
\times 10^{40}$ erg s$^{-1}$ in the $0.1-2.5\kev$ energy band) we can
compare its X-ray properties to those of
other spiral, starburst and Seyfert galaxies. Its total blue luminosity
is $L_B = 4.6 \times 10^{10}\lsun$ and the IRAS far-infrared flux between
$40\mu$m and $120\mu$m, using the expression in Helou \etal (1988), gives
a luminosity of $L_{FIR}= 6.8 \times 10^{10}\lsun$. Assuming the
conversion of Hunter \etal (1986) this corresponds to a star formation
rate for stars with masses $\geq 0.1\msun$ of about $18 \msunyr$.

From the relationship between $L_x$ and $L_{FIR}$ for starburst galaxies 
in David, Jones \& Forman (1992), and using the FIR luminosity 
of NGC\,1365 we can estimate the fraction of the X-ray luminosity 
that we might expect from the starburst component. On this basis, 
the David \etal (1992) relationship would predict a soft X-ray luminosity  
of $\sim 10^{41}\ergs$, comparable to the luminosity 
in the \rosat\ PSPC waveband (Table~\ref{tabx2}, see also 
Turner \etal 1997). We can also compare the 
values of $L_x$, $L_B$ and $L_{FIR}$ for NGC\,1365 with
those for the sample of nearby spirals and starburst galaxies in Read, Ponman
\& Strickland (1997). We find that NGC\,1365 lies roughly on the $L_x:L_B$
correlation found by Read \etal (1997), though NGC\,1365 is more X-ray luminous
than any galaxy in the sample of Read \etal (1997). These facts suggest that 
the bulk of the soft X-ray emission is due to starburst activity.

An interesting comparison can also be made between the properties of NGC\,1365
and those of WR galaxies in the survey of Stevens \& Strickland (1998). WR
galaxies are believed to be young starbursts, and we note that NGC\,1365 can be
regarded as a WR galaxy on account of the WR feature in one of the H{\small
II} regions in the central regions (region L4). However, it is unlike many
of the other WR galaxies in the sample of Stevens \& Strickland (1998), which
often are dwarf galaxies. Contini \etal (1995) included NGC\,1365 in their
sample of barred spiral WR galaxies - all of which seem to have a high
inclination. Contini
\etal (1995) suggested that starburst winds may play a role in removing
absorbing material perpendicular to the galaxy disk, making the WR
feature more likely to be observed. The X-ray spectra from the nuclear
regions of NGC\,1365 shows very little local absorption, which may provide
some support for this idea. In terms of the $L_x:L_B$ relationship,
NGC\,1365 lies rather below the trend for WR galaxies found by Stevens \&
Strickland (1998). However, the spectrum of
NGC\,1365 is very similar to those of WR galaxies. The fact that
NGC\,1365 is also a WR galaxy strengthens the contention that the soft
X-ray emission from NGC\,1365 is mostly due to starburst emission.

The relationship between radio and far-infrared luminosity extends over
many orders of magnitude and includes normal spirals, starburst galaxies
and radio-quiet Seyfert galaxies (\eg Wunderlich, Wielebinski \& Klein
1987). In the case of NGC\,1365 the far-infrared to radio lies between the
typical values for normal and starburst galaxies (Helou \etal 1985)
further indicating that the bulk of the far-infrared and radio emission is
associated with star formation processes rather than any AGN.

The Brackett\,$\gamma$ emission has been used as a predictor of X-ray emission
from starbursts (Ward 1988). In this model all of the predicted X-ray emission
from the starburst comes from massive X-ray binaries (MXRBs), which
are formed from the binary evolution of massive stars. The Brackett\,$\gamma$
emission is a consequence of recombination following ionization by the same
set of massive stars formed in the starburst. The predicted relationship
between $L_x$ (for soft X-rays) and $L_{B\gamma}$ is:

\begin{equation}
L_x (\ergs)= 7\times 10^{-35} {\frac{L_{B\gamma} L_{MXRB}}{R}}
\end{equation}
where $L_{MXRB}$ is the mean X-ray luminosity per massive X-ray binary (which
we assume to be $10^{38}\ergs$) and $R$ is the number of MXRBs per O-star. We
assume $R=500$ (Fabbiano \etal 1982). This relationship clearly does not take
into account X-ray emission from superbubbles or SNRs or an AGN.

From Forbes \& Ward (1993) we calculate that the Brackett\,$\gamma$ luminosity
in the central $6^{''}$ is $L_{B\gamma} =1.44\times 10^{39}\ergs$, which in
turn implies a total number of O-stars of around $10^5$, and an $L_x=2\times
10^{40}\ergs$. This is only a factor of 3 or so below our preferred
thermal model estimate for $L_x$ of the whole galaxy, suggesting that
the contribution of an AGN to the low energy X-ray emission from NGC\,1365 
may be small. The main caveat here is that MXRBs in our own Galaxy have much
harder spectra (typically a powerlaw with a high energy cut-off - Nagase
1989) than that observed from NGC\,1365. It is possible that the
power-law in the \asca\ observations could be a consequence of these
MXRBs. However, the line emission 
could not be generated by a population of MXRBs.

\subsection{The Location of the X-ray Nucleus}

As noted earlier we can use the optical position of the BL~Lac object
\bllac\ to register the X-ray images, which in turns enables us to
determine the centroids of the X-ray emission from the central regions of
NGC\,1365.

In Table~\ref{tabx3} we list the positions of the centroids of the X-ray
central source as determined by the PSS program, for all the \rosat\
observations. There is some scatter between the X-ray determined
positions, but this is $\leq 2^{''}$, indicating good consistency between
the results for the different \rosat\ observations. The positions of the
radio and optical nucleus are also listed in Table~\ref{tabx3}.

The mean centroid of the X-ray emission is consistent with the position
of the optical/radio nucleus. This might suggest that we are seeing
emission from a massive compact object in the nucleus. There are several 
arguments against this; first, the HRI observations (Fig.~\ref{figx4})
show that the central X-ray emission is extended and not point-like, with
an extent comparable to the radio ring (and the X-ray emission is also
centred on the radio ring). Second, the X-ray spectral properties as
seen with \rosat,
are consistent with thermal radiation, but not with power-law emission
expected from a compact object, although as the \asca\ results of Iyomoto
\etal (1997) show there is emission at higher energies from an active
nucleus. Also, a peak in the X-ray emission peaks coincident with the
position of the optical/radio nucleus is seen in other nuclear
starbursts (\cf Ptak \etal 1998).

Consequently, the X-ray emission is centred on the nucleus of the
galaxy. This does not mean that the X-ray emission is dominated by a
central point source, as the \rosat\ observations suggest that at soft
X-ray energies the emission is dominated by a nuclear starburst. While
there appears to be emission associated with an active nucleus at higher
energies its contribution at soft X-ray energies is small compared to the
hot thermal emission. {\sl AXAF}, with its $\sim 0.5^{''}$ spatial
resolution, will provide a much clearer view of the complex morphology in
the heart of NGC\,1365.

\subsection{Origin of the Radio Emission}

The spectral indices of the radio hotspots imply a non-thermal origin. The
spatial morphology - a ring around the nucleus - tells us that a compact
object is not responsible for the vast majority of the radio emission.
Supernovae or SNRs are the obvious candidates for the hotspot emission with
synchrotron emission from shock accelerated electrons the most likely emission
mechanism.

If the typical radio luminosity of a SNR is $L_{6cm} \sim 10^{33}\ergs$ 
and in the central $6^{''}$ of NGC\,1365 we have $L_{6cm} = 2.3
\times 10^{37}$ erg s$^{-1}$, this would imply 23,000 SNRs. However, the
emission from young radio supernovae can be much more luminous than this
(\ie SN1988Z; van Dyk \etal 1993), and these \lq radio supernovae\rq\
could be responsible for much of the emission from one or other of the
radio hotspots. Further to this, some \lq radio supernovae\rq\ have also
been detected as bright X-ray sources (for instance, Fabian \& Terlevich
(1996) found that SN1988Z has an X-ray luminosity in the \rosat\
waveband of $L_x\sim 10^{41}\ergs$, brighter or comparable in luminosity
to many starburst galaxies). Indeed we can correlate the positions of
the H{\small II}
regions near the centre of NGC\,1365 (Alloin \etal 1981) with the radio
hotspots. The H{\small II} region L3 and radio hotspot~A seem to be
connected, as do L12 and hotspot~E. Kristen \etal (1997), reporting on
{\sl HST} observations of NGC\,1365, find a positional coincidence between
a super star cluster and radio hotspot A. On this basis, and the probable
strong H$\alpha$ emission from this cluster, suggest that the radio
emission from hotspot A is due to a radio supernova.

It is instructive to compare $L_x/L_{6cm}$ ratios for different objects. 
Sgr A$^\ast$, the black hole at the centre of our Galaxy, has a soft X-ray
luminosity of $L_x \sim 10^{34} \ergs$ (Predehl \& Tr\"{u}mper 1994) 
and a 6cm luminosity of $L_{6cm} \sim 10^{33} \ergs$
(Marcaide \etal 1992) for an assumed distance of 8.5\,kpc. So the
$L_x/L_{6cm}$ ratio is 10. Cas A, the brightest SNR in our Galaxy, has
an $L_x/L_{6cm}$ ratio of $\sim$ 75. For radio supernovae the ratio can
be higher still ($L_x/L_{6cm}\sim$ several hundred) for SN1988Z and SN1978K.

Because of the complex extended nature of the emission, and the
limitations of the X-ray instruments it is a difficult to accurately
estimate the X-ray flux from within $6^{''}$. We estimate a value of
$\sim 1-2 \times 10^{40}\ergs$. For NGC\,1365 we have $L_{6cm} = 2.3
\times 10^{37}\ergs$, which implies a ratio of $L_x/L_{6cm}\sim 400 -1000$
within a radius of $6^{''}$. This value is rather large and could suggest
that an AGN is not the dominant energetic factor in the central regions
of NGC\,1365, and is more in line with that seen for radio supernovae.
Of relevance is the fact that the  $L_x/L_{6cm}$ ratio for NGC\,1068 is 
also in the range $500-1000$ (see Condon \etal 1982 for larger aperture
radio observations of NGC\,1068). For both NGC\,1365 and NGC\,1068 the 
emission from the nucleus (as identified at higher resolution) is much 
less than the emission from within a radio of 
$6^{''}$ (see Table~\ref{tabx4} and Muxlow \etal 1996).
We also note that large values of $L_x/L_{6cm}$ are 
possible in radio quiet QSOs (\cf Mas-Hesse \etal 1995).

\subsection{Origin of the Extended X-ray emission}

There are perhaps three main possible explanations for the extended X-ray
emission seen in NGC\,1365; first, scattering of nuclear radiation by
photoionized material, second, a large number of unresolved point
sources, and third, genuinely extended X-ray emission from a hot plasma.

\subsubsection{Scattering of Nuclear Emission}

Extended X-ray emission has also been found in the Seyfert~2 galaxy NGC\,4388
by Matt \etal (1994), using results from the \rosat\ HRI. Results from
PSPC observations of NGC\,4388 have been reported on in
Antonelli, Matt \& Piro (1997). These authors
discussed the possibility that a luminous nuclear source (\ie a supermassive
black-hole) photoionizes a large region around it, but is obscured from direct
view by a dense torus. Radiation escaping perpendicular to the torus is
scattered into the line-of-sight by electrons from the photoionized plasma.
Matt \etal (1994) discounted this idea largely on the basis of the observed
spectral properties of NGC\,4388.

In NGC\,1365, the \asca\ observations suggest the presence of a scattered
component, which might well be associated with a luminous nucleus
obscured from direct view. This scattered power-law component cannot be
responsible for the bulk of the soft emission seen in the \rosat\
waveband for the following reasons. First, the X-ray emission as seen by
the \rosat\ PSPC is very extended (8\,kpc), which would imply an extremely
luminous ionizing source, and second, and perhaps most telling, is that
the observed thermal spectrum, is too soft to result from the scattering
of a AGN continuum.

\subsubsection{Unresolved Point Sources}

Populations of stellar sources distributed throughout the bar of NGC\,1365
could mimic the extended nature of the emission by such an ensemble. These
objects could be X-ray binaries (both high and low mass), normal stars (again
both high or low mass) or supernova remnants.

Both high mass and low mass X-ray binaries have harder spectra than the
$\sim 0.7\kev$ thermal spectrum observed in NGC\,1365. The X-ray
luminosity of NGC\,1365 is rather high for a typical spiral (but not
excessively so). A model whereby a reasonably large number of MXRBs were
emitting could be plausible. We note that MXRBs are descendants of high
mass stars and will be formed in starburst events, and so this model is
not completely distinct from models associated with starbursts.

As noted in Section~4.1, the Brackett\,$\gamma$ emission can 
be used to estimate
the number of O-stars and X-ray binaries. For NGC\,1365 the number of X-ray
binaries is of order 200. If these 200 binaries were uniformly distributed in
a disk of radius 8\,kpc then the mean separation would be significantly below
that of the \rosat\ PSPC but comparable to that of the HRI. We see from both
PSPC and HRI observations that the emission from the extended component is
clearly concentrated towards the central region. Consequently such a model
with $\sim 200$ luminous MXRBs distributed through the bar of the galaxy
could account for the spatial properties seen in NGC\,1365. As we have noted
before the X-ray spectral properties would suggest that MXRBs were not the
dominant source of emission.

An origin associated with early-type stars can be quickly dismissed - these
typically have $L_x/L_{bol}\sim 10^{-7}$, implying too low a level of X-ray
emission. While magnetically active low-mass stars can have higher
$L_x/L_{bol}$, such a large population of young active stars would be
implausible.

Individual SNR can have X-ray luminosities of $L_x\sim 10^{34} -
10^{36}\ergs$, and X-ray temperatures of the right magnitude (Schlegel 1995).
Young ($\leq 20$ yr) \lq radio supernovae\rq\ can have X-ray luminosities
higher luminosities ($10^{39}\ergs$ or more), but these luminosities are short
lived. A situation where we have a large number of SNR ($10^5 - 10^6$) could
give rise to observed the X-ray luminosity. However, for any reasonable
estimate of time-scale for substantial X-ray emission to persist from an
individual SNRs (a few $10^3$ years) this would imply a very high SN rate, at
odds with observations at other wavelengths. The SN rate in the central
$6^{''}$, as inferred from the radio observations, is 0.02 SN/yr, and 0.03
SN/yr, as inferred from [FeII] emission (Forbes \& Norris 1998).

We can probably rule out a SNR origin for all of the X-ray emission by
reference to the radio observations, which imply a much smaller number of
SNRs. However, as we shall discuss below, a starburst origin of the X-ray
emission would still imply a contribution from supernovae, but in this
case the hot gas from a starburst will have been built up over a long
period of time.

We note that there have been two historical supernova in NGC\,1365; SN1957C, to
the N of the nucleus and the type Ic SN1983V, about 1 arcmin to the SW of the
nucleus (Clocchiatti \etal 1997). Neither SN are coincident with any X-ray or
radio source.

\subsubsection{Extended hot gas}

The most plausible physical origin for extended hot gas in the disk of
NGC\,1365 is from a starburst. Here a burst of star-formation results in the
production of large numbers of massive OB stars, which via their stellar winds
and subsequent supernova inflate X-ray emitting superbubbles, which will
eventually merge and blow-out to form a superwind, such as seen in M82
(Strickland, Ponman \& Stevens 1997, Moran \& Lehnert 1997).

The simple analysis concerning the Brackett\,$\gamma$ emission suggests the
presence of around $10^5$ O9 stars (or their equivalent). The mass-loss rate
of such a star is $\sim 10^{-6}\msunyr$ and the wind kinetic energy is of
order $10^{36} \ergs$, so the total kinetic energy from such a cluster of
stars (assuming they are all concentrated towards the centre of the 
galaxy) will
be $10^{41}\ergs$. As only a fraction of the kinetic energy of the stellar
winds can be converted into X-ray emission such an assembly of stars would
not, by themselves, be able to power the X-ray emission from NGC\,1365.
However, this analysis does not account for the contribution from SNRs and
X-ray binaries, and the hot stars could make a significant contribution to the
X-ray emission via their stellar winds.

\subsection{Comparison with other Seyfert~2 galaxies}

In order to understand the radio/X-ray properties of NGC\,1365 we now briefly
discuss similar observations of related galaxies.

Turner \etal (1993), in addition to presenting results on NGC\,1365, also
presented \rosat\ PSPC results on several Seyfert~2 galaxies. Of note is
that it was not possible to obtain a good fit with a power-law model for
NGC\,1365, while much better fits were obtained for the other galaxies in
the sample. Also, the other galaxies were typically an order of magnitude
more X-ray luminous than NGC\,1365, and were much more overluminous
compared to their blue luminosity. This would suggest that NGC\,1365 is
not a typical Seyfert~2, but as we shall see in the discussion below,
there are some striking similarities between the X-ray properties of
NGC\,1365 and some other galaxies classified as Seyfert~2's.

Awaki (1997) reviews the X-ray properties of Seyfert~2s, finding an
extremely wide range of luminosities, with some galaxies having
luminosities of $5\times 10^{43}\ergs$, while others have $L_x\sim
10^{41}\ergs$. The radio emission from Seyfert~2s may cover an even
larger range of $4-5$ orders of magnitude (see for example Simpson \etal
1996).

Wilson \etal (1992) have reported on \rosat\ HRI observations of the canonical
Seyfert~2 galaxy NGC\,1068. These observations revealed a compact nuclear
source, circumnuclear emission (with a diameter of 1.5\,kpc) and much more
extended emission (diameter 13\,kpc) associated with the starburst disk. Wilson
\etal (1992) concluded that the nuclear and circumnuclear X-ray emission were
from thermal gas associated with an outflowing wind.

Higher spectral resolution X-ray observations of NGC\,1068 with both 
{\it BBXRT} and \asca\ (Marshall \etal 1993; Ueno \etal 1994) found
strong line emission from a range of different species, including strong
Fe-K emission. In addition, the X-ray continuum could be fitted with a
model with two soft thermal components (with one component with $\kt<
0.2\kev$ and a 2nd component with $\kt\sim 0.4-0.7\kev$) and a power-law
component, with a photon index of $\sim 1.3$. Importantly, the power-law
component does not appear to be highly absorbed. According to Marshall
\etal (1993) the 2-10\kev luminosity of NGC\,1068 (dominated by the
power-law component) is $\sim 2\times 10^{41}\ergs$, approximately an
order of magnitude brighter than the powerlaw component for NGC\,1365 
(Iyomoto \etal 1997).

Marshall \etal (1993) and Ueno \etal (1994) developed a model for NGC\,1068
whereby the X-ray emission is due to both starburst emission and
scattered emission from a much more luminous (possibly $10^{43}\ergs$)
active nucleus. In this picture we see no direct X-ray emission from the
active nucleus (it being obscured by material with an extremely high
column), only emission scattered into the line of sight by both warm and
hot absorbing material.

It is worth noting that at radio wavelengths the nucleus of NGC\,1068 is much
brighter than NGC\,1365 (Muxlow \etal 1996). The source that Muxlow \etal
(1996) identify with the nucleus of NGC\,1068 in addition to having 
an inverted spectrum, is well over an order of magnitude brighter than
the source identified with the nucleus in NGC\,1365. With larger 
apertures NGC\,1068 is still much brighter at radio wavelengths 
than NGC\,1365 (\cf Section 4.3; Condon \etal 1982). On 
spatial scales of several arcseconds the radio emission is dominated by 
starburst/SNR emission, with only a small contribution from the nucleus.

Matt \etal (1994) have reported on \rosat\ HRI observations of NGC\,4388, a
highly inclined spiral galaxy (SB(s)b pec) near the core of the Virgo cluster,
finding extended emission with a radius of 4.5\,kpc. Matt \etal (1994) did not
find any evidence of a central source, constraining the emission from such a
point source to less than 20\% of the total.

NGC\,4388 also shows weak broad H$\alpha$ emission in extended off nuclear
regions (Shields \& Filippenko 1988), which has been interpreted as being
due to scattered light from an obscured nucleus (as per the standard
unified model of Seyfert~2 nuclei). NGC\,4388 also has two extended and
asymmetric ionization cones. Iwasawa \etal (1997) have also studied the
X-ray properties of NGC\,4388 including \asca\ observations. The \asca\
observations reveal the presence of a strong and highly absorbed
($\nh=4\times 10^{23}\cmn$) power-law component coming from the nucleus,
along with narrow Fe line emission at $6.4\kev$ (though this emission
line has a large equivalent width (EW$\sim 500$eV - Iwasawa \etal 1997).

Brandt, Halpern \& Iwasawa (1996) have reported on \rosat\ PSPC and HRI
observations of the composite starburst/Seyfert~2 NGC\,1672 (SB(s)b). Of three
X-ray sources associated with NGC\,1672, the nuclear source (X-1) has an
unabsorbed luminosity of $L_x \sim 10^{40}\ergs$ in the \rosat\
waveband, $\kt=0.7\kev$ and a low absorbing column 
($\nh=1.6\times 10^{20}\cmn$). The X-ray luminosity of
NGC\,1672 is somewhat lower than for NGC\,1365, but otherwise the X-ray
properties are remarkably comparable. NGC\,1672 also shows a compact radio
source located at the optical nucleus, with an almost circular ring of
emission, as seen in NGC\,1365 (Sandqvist \etal 1995). Consequently, NGC\,1672
may indeed be very similar to NGC\,1365, and the Seyfert nucleus may also not
play a dominant role in this galaxy.

\subsection{Is there a Seyfert Nucleus in NGC\,1365 ?}

As noted in the introduction, NGC\,1365 has been classified as a galaxy
with a Seyfert nucleus on the basis of both broad and narrow H$\alpha$
emission. On the basis of the \rosat\ data there is no real evidence at
soft X-ray energies of a Seyfert type nucleus in NGC\,1365. However, 
the \asca\ data
does suggest the presence of an active nucleus. We suggest 
that the active nucleus in NGC\,1365 is a low-luminosity analogue 
of that in NGC\,1068, with a luminosity of 5-10\% of that in NGC\,1068.
So that, while in NGC\,1068 it has been suggested that
the AGN perhaps provides approximately 50\% of the 
bolometric luminosity (Telesco \etal 1984). We suggest 
that in NGC\,1365 the fraction provided by the AGN will be 
correspondingly smaller.

There is also evidence of something resembling an active nucleus at radio
wavelengths, but whether it is strong enough to warrant NGC\,1365 being
classified as a Seyfert nucleus is questionable. At optical wavelengths,
there is some evidence of AGN related activity - with an outflow cone
seen in [O{\small III}] $\lambda 5007$, though an outflow like this could
be generated by starburst activity. As to the broad H$\alpha$ emission -
could the radio supernova associated with radio hotspot A be responsible
for this?

Recently, when discussing massive black holes in galaxies it has been
become popular to suggest that many black holes are not accreting at the
Eddington rate but rather the much reduced rate associated with an
advection dominated accretion flow (ADAF; Narayan 1996).  We consider the
presence of an ADAF in NGC\,1365 to be unlikely, on the basis of both the
radio and X-ray spectra. Such ADAFs predict a slightly positive spectral
index at radio wavelengths. We find that the 6/3 cm index is quite steep
at $-1.1$, and we do not see much evidence for a very hot bremsstrahlung
spectrum from the nucleus (though this component could possibly be
mistaken for the observed power-law component), though its absence would
have to be confirmed by {\sl AXAF}. An unobscured ADAF would also not
produce  strong Fe-K line emission (Narayan 1996).

NGC\,1365 is a highly inclined system, with $i\sim 60^\circ$, so that, in a
more speculative vein, we could be looking down onto a superwind, such as seen
in M82. Neither the X-ray temperature nor luminosity are unreasonable for such
a model, and this idea merits further investigation. As mentioned earlier, an
outflow cone has been observed in this galaxy, though it was thought to be
related more to AGN activity.

In summary, we have presented radio and X-ray observations of the
prominent barred spiral NGC\,1365. On the basis of soft X-ray and radio
observations we find little if any evidence of an AGN, but much more
evidence for strong star-forming activity. However, 
strong Fe-K emission does suggest 
an active nucleus. Comparison with other Seyfert galaxies,
such as NGC\,1068, suggest that any AGN in NGC\,1365 is not energetically
dominant at most wavelengths. Consequently, we suggest that the 
active nucleus in NGC\,1365 is a low luminosity
analogue of the nucleus in NGC\,1068. Even in the 
nucleus of NGC\,1365, AGN activity is accompanied by very significant
starburst emission, and  outside of the nuclear region starburst
activity completely dominates.

\section*{Acknowledgements}

The anonymous referee is thanked for a very helpful report which
significantly improved the quality of the paper.
 
IRS acknowledges funding from PPARC. The data analysis presented in this
paper made use of the {\em STARLINK} node at the University of Birmingham, and
the {\em ASTERIX} package. This research has made use of data 
obtained from the Leicester Database and Archive Service at
the Department of Physics and Astronomy, Leicester University, UK.
The Digitised Sky Surveys were produced at the Space Telescope Science 
Institute under U.S. Government grant NAG W-2166. The images of
these surveys are based on photographic data obtained using the Oschin
Schmidt Telescope on Palomar Mountain and the UK Schmidt Telescope.


\begin{thebibliography}{99}

\bibitem{}Allan D.\,J., 1993, ASTERIX User Note No. 4.

\bibitem{}Alloin D., Edmunds M.\,G., Lindblad P.\,O., Pagel B.\,E.\,J., 1981,
A\&A, 101, 377 
 
\bibitem{}Antonelli L.\,A., Matt G., Piro L., 1997, A\&A, 317, 686

\bibitem{}Awaki H., Ueno S., Koyama K., Iwasawa K., Kunieda H., 1997, Adv.
Sp. Res., 19, 95

\bibitem{}Brandt W.\,N., Halpern J.\,P., Iwasawa K., 1996, MNRAS, 281, 687

\bibitem{}Briel U.\,G., \etal 1995, \lq\lq The \rosat\ Users'
Handbook\rq\rq\ (Garching: MPE). 


\bibitem{}Clocchiatti A., \etal 1997, ApJ, 483, 675

\bibitem{}Condon J.\,J., Condon M.\,A., Gisler G., Puschell J.\,J., 
1982, ApJ, 252, 102

\bibitem{}Condon J.\,J., Yin Q.\,F., 1990, ApJ, 357, 97

\bibitem{}Conti P.\,S., 1991, ApJ, 377, 115

\bibitem{}Contini T., Davoust E., Consid\`{e}re S., 1995, A\&A, 303, 440

\bibitem{}David L.\,P., Jones C., Forman W., 1992, ApJ, 388, 92

\bibitem{}Fabbiano G., Feigelson E., Zamorani G., 1982, ApJ, 256, 397

\bibitem{}Fabbiano G., Kim D.-W., Trinchieri G., 1992, ApJS, 80, 531

\bibitem{}Fabian A.\,C., Terlevich R.\,J., 1996, MNRAS, 280, L5

\bibitem{}Forbes D.\,A., Norris R.\,P., 1998, MNRAS, 300, 757

\bibitem{}Forbes D.\,A., Ward M.\,J., 1993, ApJ, 416, 150

\bibitem{}Ghosh S.\,K., Verma R.\,P., Rengarajan T.\,N., Das B., Saraiya
H.\,T., 1993, ApJS, 86, 401

\bibitem{}Helou G., Khan I.\,R., Malek L., Boehmer L., 1988, ApJS, 68, 151

\bibitem{}Hjelm M., Lindblad P.\,O., 1996, A\&A, 305, 727

\bibitem{}Hunter D.\,A., Gillet F.\,C., Gallagher J.\,S., Rice W.\,L., 
Low F.\,J. 1986, ApJ, 303, 171

\bibitem{}Irwin M., 1992, Gemini, 37, 1

\bibitem{}Iwasawa K., Fabian A.\,C., Ueno S., Awaki H., Fukazawa Y.,
Matsushita K., Makishima K., 1997, MNRAS, 285, 683

\bibitem{}Iyomoto N., Makishima K., Fukazawa Y., Tashiro M., Ishisaki Y., 
1997, PASJ, 49, 425

\bibitem{}Komossa St., Schulz H., 1998, A\&A, 339, 345

\bibitem{}Kristen H., J\"{o}rs\"{a}ter S., Lindblad P.\,O., Boksenberg A.,
1997, A\&A, 328, 483

\bibitem{}Lamer G., Brunner H., Staubert R.,  1996, A\&A, 311, 384

\bibitem{}Maccacaro T., Wolter A., McLean B., Gioia I.\,M., Stocke J.\,T., 
Della Ceca R., Burg R., Faccini R., 1994, Astrophys. Lett., 29, 267

\bibitem{}Maiolino R., Rieke G.\,H., 1995, ApJ, 454, 95

\bibitem{}Marcaide J.\,M., \etal 1992, A\&A, 258, 295

\bibitem{}Marshall F.\,E., \etal 1993, ApJ, 405, 168

\bibitem{}Mas-Hesse J.\,M., Rodriguez-Pascual P.\,M., Sanz Fern\'{a}ndez de
C\'{o}rdoba L., Mirabel I.\,F., Wamsteker W., Makino F., Otani C., 1995, A\&A,
298, 22

\bibitem{}Matt G., Piro L., Antonelli L.\,A., Fink H.\,H., Meurs E.\,J.\,A., 
Perola G.\,C., 1994, A\&A, 292, 13

\bibitem{}Moran E.\,C., Lehnert M.\,D., 1997, ApJ, 478, 178

\bibitem{}Muxlow T.\,W.\,B., Pedlar A., Holloway A.\,J., 
Gallimore J.\,F., Antonucci R.\,R.\,J., 1996, MNRAS, 278, 854

\bibitem{}Nagase F., 1989, PASJ, 41, 1

\bibitem{}Narayan R., 1996, ApJ, 462, 136

\bibitem{}Phillips A.\,C., Conti P.\,S., 1992, ApJ, 395, L91

\bibitem{}Predehl P., Tr\"{u}mper J., 1994, A\&A, 290, L29

\bibitem{}Ptak A., Serlemitsos P., Yaqoob T., Mushotzky R., 1998, ApJ, (in
press)

\bibitem{}Read A.\,M., Ponman T.\,J., Strickland D.\,K., 1997, MNRAS, 286, 626

\bibitem{}Saikia D.\,J., Pedlar A., Unger S.\,W., Axon, D.\,J.,
1994, A\&A, 270, 46

\bibitem{}Sandage A, Tammann G.\,A., 1981, A Revised
Shapley-Ames Catalogue of Bright Galaxies. Carnegie Inst. of Washington.

\bibitem{}Sandqvist Aa, J\"{o}rs\"{a}ter S., Lindblad P.\,O.,
1995, A\&A, 295, 585

\bibitem{}Schlegel E.\,M., 1995, Rep. Prog. Phys., 58, 1375

\bibitem{}Shields J.\,C., Filippenko A.\,V., 1988, ApJ, 332, 55

\bibitem{}Simpson C., Forbes D.\,A., Baker A.\,C., Ward M.\,J., 1996, 
MNRAS, 283, 777

\bibitem{}Stark A.\,A., Gammie C.\,F., Wilson R.\,W., Bally J., 
Linke R.\,A., Heiles C., Hurwitz M., 1992, ApJS, 79, 77

\bibitem{}Stevens I.\,R., Strickland D.\,K., 1998, MNRAS, 294, 523

\bibitem{}Strickland D.\,K., Ponman T.\,J., Stevens I.\,R., 1997, A\&A, 
320, 378

\bibitem{}Strickland D.\,K., Stevens I.\,R., 1998, MNRAS, 297, 747

\bibitem{}Telesco C.\,M., Becklin E.\,E., Wynn-Williams C.\,G., 
Harper D.\,A., 1984, 282, 427

\bibitem{}Turner T.\,J., Urry C.\,M., Mushotzky R.\,F., 1993, ApJ, 418, 653

\bibitem{}Turner T.\,J., George I.\,M., Nandra K., Mushotzky R.\,F., 1997, 
ApJS, 113, 23

\bibitem{}Ueno S., Mushotzky R.\,F., Koyama K., Iwasawa K., Awaki H., 
Hayashi I., 1994, PASJ, 46, L71

\bibitem{}van Dyk S.\,D., Weiler K.\,W., Sramek R.\,A., Panagia N.,
1993, ApJ, 419, L69

\bibitem{}V\'{e}ron P., Lindblad P.\,O., Zuiderwijk E.\,J., V\'{e}ron M.\,P., 
Adam G., 1980, A\&A, 87, 245

\bibitem{}Ward M.\,J., 1988, MNRAS, 231, 1P

\bibitem{}Wilson A.\,S., Elvis M., Lawrence A., Bland-Hawthorn J.,
1992, ApJ, 391, L75

\bibitem{}Wunderlich E., Wielebinski R., Klein U., 1987, A\&AS, 69, 487

\bibitem{}Xue S.-J., Otani C., Mihara T., Cappi M., Matsuoka M., 1998,
PASJ, 50, 519

\end{thebibliography}
\end{document}